\pdfoutput=1

\documentclass[twocolumn]{aastex7}
\usepackage{CJK}
\usepackage{enumitem}
\usepackage{amsmath}

\newcommand{\msun}{M_\odot}

\newcommand{\mj}{M_{\rm J}}

\newcommand{\kms}{{\rm km} \ {\rm s}^{-1}}

\defcitealias{hamer2019hot}{HS19}

\begin{document}
\begin{CJK*}{UTF8}{gbsn}

\title{Are Hot Jupiters Tidally Disrupted During Stellar Main Sequence?}

\author[0009-0007-9015-9451]{Qingru Hu (胡清茹)}
\affiliation{Department of Astronomy, Tsinghua University, Beijing 100084, China}
\email{huqr24@mails.tsinghua.edu.cn}

\author[0000-0003-4027-4711]{Wei Zhu (祝伟)}
\affiliation{Department of Astronomy, Tsinghua University, Beijing 100084, China}
\email{weizhu@tsinghua.edu.cn}

\author[0000-0003-3250-2876]{Yang Huang (黄样)}
\affiliation{School of Astronomy and Space Science, University of Chinese Academy of Sciences, Beijing 100049, China}
\affiliation{National Astronomical Observatories, Chinese Academy of Sciences, Beijing 100101, China}
\email{huangyang@ucas.ac.cn}

\author[0009-0008-7479-0742]{Bowen Zhang (张博闻)}
\affiliation{School of Astronomy and Space Science, University of Chinese Academy of Sciences, Beijing 100049, China}
\affiliation{National Astronomical Observatories, Chinese Academy of Sciences, Beijing 100101, China}
\email{zhangbw@bao.ac.cn}

\correspondingauthor{Qingru Hu, Wei Zhu}
\email{huqr24@mails.tsinghua.edu.cn, weizhu@tsinghua.edu.cn}

\begin{abstract}
Once hot Jupiters (HJs) reach their very close orbits, they are expected to experience orbital decay due to tidal interactions with their host star. However, the strength of tidal dissipation is highly uncertain, and it remains an open question whether HJs are tidally disrupted during the stellar main sequence.
A previous study found that HJ hosts have a smaller Galactic total velocity dispersion than their field star counterparts, which they interpreted as evidence of tidal disruption. We revisit this study and find that, after using the more reliable vertical velocity dispersion ($\sigma_W$) as the age indicator and accounting for the heterogeneity and anisotropy of their HJ sample, the kinematic age difference between their HJ hosts and matched field stars is significantly reduced.
As an independent check, we collect HJs newly discovered by TESS and find that their $\sigma_W$ is statistically similar to that of matched field stars.
We also find no statistically significant $\sigma_W$ difference between the field stars and the theoretically vulnerable ultra-hot Jupiters with $P<2$ d.
Our results suggest that, after accounting for systematics in the age--velocity dispersion relation, there is no statistically strong evidence from the stellar kinematics that a large fraction of hot Jupiters around Sun-like stars are tidally destroyed during the stellar main sequence.
\end{abstract}


\section{Introduction}\label{sect:intro}

The discovery of hot Jupiters (HJs) and their unusual properties pose significant challenges to classical planet formation theories derived from our Solar System \citep{mayor1995jupiter}. To explain the origin of these planets, several models have been proposed, including in situ formation \citep[e.g.,][]{batygin2016situ,boley2016situ}, disk migration \citep[e.g.,][]{lin1996orbital}, and high-eccentricity migration \citep[e.g.,][]{rasio1996dynamical,wu2003planet}. However, no single model can satisfactorily account for all the observational evidence, and the formation pathway of hot Jupiters remains an open question \citep[][]{dawson2018origins}. 

The future of hot Jupiters is also uncertain. Once a HJ reaches its very close-in orbit, tides raised on its host star are expected to shrink the orbit of the HJ by transferring angular momentum from the planetary orbit to the stellar spin \citep[e.g.,][]{rasio1996tidal,jackson2008tidal}.
Given long enough time, all HJs will eventually be engulfed by their host stars since their orbital angular momentum is generally too low to synchronize the stellar spin \citep{hut1980stability}.
Nevertheless, the timescale of such tidal decay, or equivalently the (effective) ``modified tidal quality factor" of the star, $Q'_*$, defined as the quality factor $Q_\star$ divided by $2/3$ of the Love number $k_2$ \citep{goldreich1966solar}, remains highly uncertain due to our poor understanding of the tidal dissipation mechanisms within stars. Commonly assumed to be a fixed value in observations, the modified tidal quality factor actually depends on the mass and orbital period of the planet as well as the interior structure of the star \citep[e.g.,][]{barker2010internal,lai2012tidal,ogilvie2014tidal,essick2015orbital}.

The most straightforward way to understand the tides-driven orbital decay is to measure the orbital decay rate of individual hot Jupiters. Fortunately and unfortunately, after many years of searching \citep[e.g.,][]{bouma2019wasp,patra2020continuing,wang2024long,winn2025orbital}, WASP-12 b stands out as the only one with unambiguous evidence of tidal inspiral \citep{maciejewski2016departure,patra2017apparently,yee2020orbit}. According to \citet{yee2020orbit}, the orbital decay timescale of WASP-12 b is $\sim3$ Myr (corresponding to $Q'_*\sim2\times10^5$), much shorter than the age $\sim3$ Gyr of its host star \citep{hebb2009wasp}. It could be naturally explained if WASP-12 has become a subgiant star, as the nonlinear wave-breaking of the dynamical tide within the radiative core could lead to efficient dissipation \citep{weinberg2017tidal,barker2020tidal}.
However, whether the host star of WASP-12 b has evolved off the main sequence remains debated \citep{bailey2019understanding}. If WASP-12 is indeed still a main-sequence star, current tidal theories do not provide a viable mechanism to account for such strong tidal interactions \citep{chernov2017dynamical,weinberg2024orbital,sun2025numerical}, unless an as-yet-undetected external perturber is driving the obliquity tides \citep{millholland2018obliquity}.

An alternative way to study the tidal-driven orbital evolution of hot Jupiters is through statistical analyses of a large sample.
For example, some previous studies found that Sun-like stars hosting HJs rotate faster than comparable stars without such planets \citep[e.g.,][]{pont2009empirical,husnoo2012observational,penev2018empirical,tejadaarevalo2021further} or have smaller gyrochronological age estimates than isochronal ones \citep[e.g.,][]{brown2014discrepancies,maxted2015comparison}. In addition, it was proposed that the semi-major axis distribution of HJs show evidence of orbital decay and tidal engulfment \citep[e.g.,][]{jackson2009observational,cameron2018hierarchical,millholland2025empirical,lazovik2026observational}. 

If a non-negligible fraction of HJs are tidally disrupted during the stellar main sequence, another natural consequence is that stars hosting HJs should be statistically younger than their field star counterparts.
\citet[hereafter HS19]{hamer2019hot} used the Galactic total velocity dispersion of stars as a proxy for their age and claimed that the hot Jupiter hosts were kinematically younger than their matched field stars as well as the hosts of longer-period Jupiters, thus supporting the idea that hot Jupiters undergo substantial orbital decay during the stellar main sequence. This result has been widely received by the community \citep[e.g.,][]{chen2023evolution,miyazaki2023evidence,banerjee2024hoststar}. Based on it, several recent papers further tried to explain the hot Jupiter occurrence rate evolution via a combination of early and late arrivals and put constraints on the dynamical evolution of hot Jupiters \citep{chen2026origin, schmidt2026most}.

\begin{figure*}[t]
    \centering
    \includegraphics[width=\linewidth]{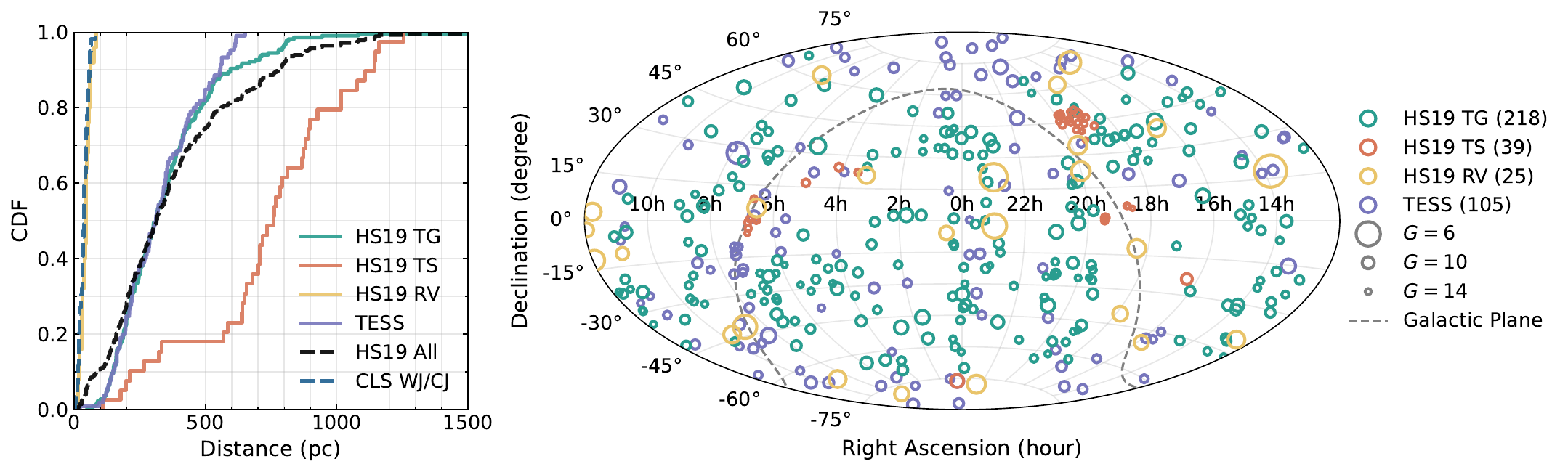}
    \caption{\textit{Left}: Cumulative distribution functions of host star distances for different hot Jupiter (HJ) and warm/cold Jupiter (WJ/CJ) samples. The green, red, and yellow lines represent the distance distributions of HJ hosts discovered by ground-based transit (TG), space-based transit (TS), and radial velocity (RV), respectively, from the updated \citetalias{hamer2019hot} sample. The purple solid line corresponds to the TESS HJ sample. The black dashed line shows the distance distribution for all HJ hosts in the updated \citetalias{hamer2019hot} sample, and the blue dashed line shows that for the WJ/CJ sample from the California Legacy Survey (CLS). \textit{Right}: Spatial distribution of hot Jupiter host stars on the celestial sphere. The point size scales with the apparent Gaia $G$ magnitude of the host stars. Green, red, and yellow circles indicate HJ hosts discovered by ground-based transit (TG), space-based transit (TS), and radial velocity (RV), respectively, from the updated \citetalias{hamer2019hot} sample. Purple circles represent HJ hosts discovered by TESS. The gray dashed line marks the Galactic plane.}
    \label{fig:sky-dist}
\end{figure*}

\begin{figure}
    \centering
    \includegraphics[width=\linewidth]{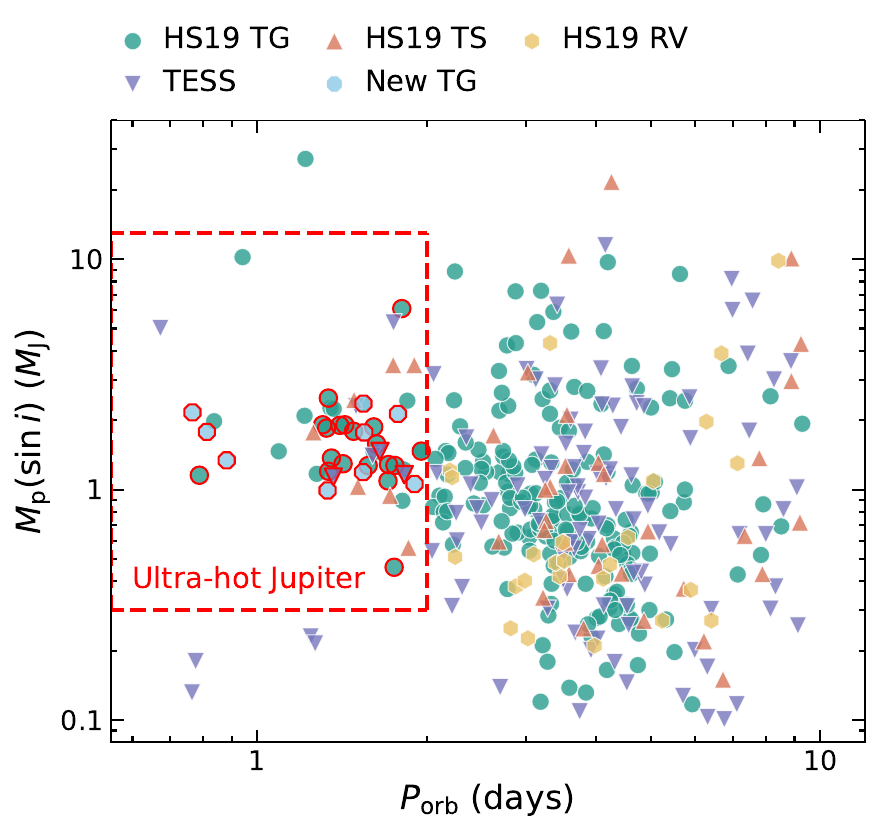}
    \caption{
    Planetary (minimum) mass-period diagram for the updated \citetalias{hamer2019hot} sample (HS19 TG, HS19 TS, and HS19 RV), the TESS HJ sample (TESS), and hot Jupiters newly discovered by ground-based transit (New TG). Ultra‑hot Jupiters are defined as those with orbital period $P_{\rm orb} < 2$ d and planetary (minimum) mass $0.3 < M_{\rm p}(\sin i) < 13\ \mj$, as indicated by the red dashed box. The ultra-hot Jupiters used in our analysis in Section~\ref{sect:ultra-hot} are marked out by bold red edges.}
    \label{fig:m-p}
\end{figure}

\citetalias{hamer2019hot} and the later studies adopted the standard definition of hot Jupiters from observations: a giant planet with orbital period $P<10$ days. From the theoretical perspective, the tidal decay timescale is a steep function of the orbital period. According to the ``constant phase lag" model of \citet{goldreich1966solar}, this timescale is
\begin{equation} \label{eqn:timescale}
\begin{aligned}
    &\tau_{\rm decay} = \frac{4}{27} G^{\frac53} \frac{Q_*' M_*^{\frac{8}{3}}}{R_*^5 M_{\rm p}} \left( \frac{P}{2\pi} \right)^{\frac{13}{3}} \\
    &\approx 10 \ {\rm Gyr} \left(\frac{Q_*'}{10^5}\right) \left(\frac{P}{5 \ {\rm d}}\right)^{\frac{13}{3}}  \left(\frac{M_*}{\msun}\right)^{\frac{8}{3}} \left(\frac{R_*}{R_\odot}\right)^{-5} \left(\frac{M_{\rm p}}{\mj}\right)^{-1},
\end{aligned}
\end{equation}
where $R_*$ is the stellar radius, and $M_*$ and $M_{\rm p}$ are the mass of the star and the hot Jupiter, respectively. Assuming a Sun-like star and a strong tide with $Q_*'\approx 10^5$, only hot Jupiters with $P<5$ days will be tidally disrupted within the main sequence stage of $\sim 10$ Gyr. For main-sequence stars with more typical values, theories suggest that tides can only be strong enough to drive substantial orbital decay of Jupiter-mass planets with $P\lesssim 2\,$d \citep[e.g.,][]{weinberg2024orbital}. Such planets contribute a minor fraction of all hot Jupiters in \citetalias{hamer2019hot}.

We are therefore motivated to revisit the \citetalias{hamer2019hot} study to understand whether the hot Jupiter hosts are indeed kinematically younger than the field stars of similar properties. There are a few issues in the statistical analysis of that study. First, although the stellar age--velocity dispersion relation \citep[AVR,][]{stromger1946, wielen1977} has been widely used to connect the stellar kinematics with age, this relation is more valid for the dispersion in vertical velocity, namely the velocity component along the vertical direction to the Galactic disk \citep[e.g.,][]{seabroke2007revisiting, aumer2016, ting2019, chen2021planets}, rather than in the total velocity that is used in \citetalias{hamer2019hot}. A closer check also reveals that \citetalias{hamer2019hot} adopted an unconventional (or erroneous) definition of velocity dispersion.
\footnote{See their Equation~(1), and we confirm that it is not due to typographical errors (Hamer \& Schlaufman, private comm.). The same definition was also probably used in \citet{hamer2020ultrashortperiod,hamer2022evidence,hamer2024kepler,schmidt2024resonant,schmidt2026most}.}
As will be shown later, adopting the vertical velocity dispersion rather than the total velocity dispersion reduces the statistical significance in the velocity dispersion difference (see also Fig.~S2 of \citealt{chen2023evolution}), although it alone cannot account for the velocity dispersion difference completely.

A more important issue is the heterogeneity of their hot Jupiter sample. \citetalias{hamer2019hot} used hot Jupiters discovered via different detection methods and by various surveys/missions, which encountered different sensitivity limits. As shown in Fig.~\ref{fig:sky-dist}, hot Jupiters from ground-based radial velocity (RV) surveys are all within 100\,pc, whereas the typical distance to hot Jupiters from ground-based transit surveys, which contribute the majority of the hot Jupiters in the \citetalias{hamer2019hot} sample, is a few hundred pc. Hot Jupiters from the space-based transit surveys, which in \citetalias{hamer2019hot} meant CoRoT, Kepler, and K2, can be farther than 1\,kpc, and their locations on the sky plane are highly clustered. The AVR is known to vary across such a large spatial volume \citep[e.g.,][]{sharma2021fundamental, sun2025age}, and in no way was the anisotropy in the spatial distribution of hot Jupiter hosts accounted for when \citetalias{hamer2019hot} constructed the field star comparison sample. As shown later, after taking into account the hot Jupiter sample heterogeneity and anisotropy, we are able to explain the observed velocity dispersion difference between hot Jupiter hosts and the field stars.

The paper is organized as follows: in Section~\ref{sect:HS19}, we take into account the aforementioned issues and revisit the \citetalias{hamer2019hot} hot Jupiter sample with minimum modifications; in Section~\ref{sect:tess}, we apply the same method to the newly discovered hot Jupiter sample from the Transiting Exoplanet Survey Satellite (TESS, \citealt{ricker2015transiting}), which serves as an updated and independent check; motivated by the theoretical argument (Equation~\ref{eqn:timescale}), we apply our method to the sample of ultra-hot Jupiters with $P<2$ days in Section~\ref{sect:ultra-hot}. A comparison to the warm/cold Jupiter systems is shown in Section~\ref{sec:wcj}. Our results are summarized in Section~\ref{sect:summary}.

\section{Revisiting HS19 Sample}\label{sect:HS19}

\subsection{Sample Reconstruction}\label{subsect:sample-construct}

We start from the list of 338 hot Jupiter hosts presented in Table~1 of \citetalias{hamer2019hot} and cross match with NASA Exoplanet Archive \citep[NEA,][]{neatable} \footnote{The NASA Exoplanet Archive table was downloaded on January 18, 2026.}
to collect the stellar and planetary parameters. \citetalias{hamer2019hot} defined hot Jupiter to be planet with orbital period $P<10$ days and planetary mass $M_{\rm p} (\sin i) >0.1 \mj$.
\footnote{The symbol $M_{\rm p} (\sin i)$ corresponds to the column \texttt{pl\_bmassj} in the NASA Exoplanet Archive, which records the best planet mass estimate available, in order of preference: $M_{\rm p}$, $M_{\rm p} \sin i / \sin i$, or $M_{\rm p} \sin i$.}
With the NEA parameters, we find that 11 systems in the \citetalias{hamer2019hot} sample no longer host hot Jupiters according to their fiducial definition. 
Additionally, we exclude 15 systems from our sample to ensure the homogeneity of the hot Jupiter sample: three are excluded because they were discovered by neither transit or RV, seven are excluded because the hosts have relatively close stellar companion(s),
and five are excluded because the hosts are in globular/open clusters.

We then query the Gaia Data Release 3 \citep[Gaia DR3;][]{2023gaia} data for the astrometric parameters of the hot Jupiter hosts. After applying the astrometry cuts 1--3 in Appendix~\ref{appen:cuts}, We further exclude two systems from our sample. For hot Jupiter hosts without RVs from NEA, we use the RV measurement from Gaia DR3.

We follow \citetalias{hamer2019hot} to remove stars that are not main sequence. Since our sources are all very nearby, the Gaia extinction corrections are reliable; thus, we correct for extinction and reddening of the hot Jupiter hosts using extinction (\texttt{ag\_gspphot}) and reddening (\texttt{ebpminrp\_gspphot}) from Gaia-DR3 GSP-Phot Aeneas best library \citep{andrae2023gaia}. For stars without Gaia DR3 extinction or reddening measurements, we calculate reddening $E(B-V)$ using the 3D dust map of \texttt{Bayestar19} \citep{green20193d} implemented in the Python package \texttt{dustmaps} \citep{green2018dustmaps}, and convert it to Gaia extinction $A_G$ and reddening $E(\rm{BP-RP})$ using the extinction coefficients in \citet{casagrande2018use}. The 3D dust map fails for some hot Jupiter hosts that are very close, and in such cases we simply adopt zero extinction.

With the stellar positions on the intrinsic color vs.\ absolute magnitude diagram, we reject hot Jupiter hosts that lie more than one magnitude above the Pleiades main-sequence relation of \citetalias{hamer2019hot}. We further restrict our analysis to FGK-type stars with $0.5\leq (\rm{BP-RP})_{\rm 0} \leq 1.5$ in order to exclude the very early/late type but rare objects \citep{pecaut2013intrinsic}. A number of 26 hot Jupiter hosts are excluded according to the main-sequence and stellar type cuts.
Finally, we also exclude two hot Jupiter hosts (ups And and POTS-1) for their extreme small/large Gaia $G$ magnitudes, which results in too few/many field stars that can be matched according to our criteria in Section~\ref{subsect:HS19-res}.

After all these necessary clean up steps, we are left with 282 hot Jupiter hosts in the updated \citetalias{hamer2019hot} sample, whose distance, equatorial coordinates, and the Gaia $G$ magnitude are plotted in Fig.~\ref{fig:sky-dist}. Depending on the discovery methods and facilities, we further label the stars as “HS19 TG”, “HS19 TS”, and “HS19 RV”, which correspond to hot Jupiters discovered via ground-based transit, space-based transit, and RV, respectively. These three subsamples contain 218, 39, and 25 stars, respectively. We show in Fig.~\ref{fig:m-p} the (minimum) mass vs.\ orbital period of the hot Jupiters from the three subsamples. The identifiers of our updated \citetalias{hamer2019hot} HJ sample is listed in Appendix~\ref{sect:target}.

We use \texttt{gal\_uvw} in PyAstronomy\footnote{https://github.com/sczesla/PyAstronomy} to calculate the Galactic rectangular velocities of the stars relative to the local standard of rest (LSR) ($U,V,W$). We have adopted the solar peculiar motion of $(U_\odot, V_\odot, W_\odot)=(11.1,12.24,7.25) \ {\rm km} \ {\rm s}^{-1}$ from \citet{schonrich2010local}. The derived dispersion is insensitive to the value of this peculiar motion, although it is not precisely determined \citep[e.g.,][]{huang2015lsr,wang2021lsr}.

\subsection{Field Star Sample}
We construct a comparison sample of field stars without known hot Jupiters from Gaia DR3, selecting objects with high-quality five-parameter astrometric solutions and radial velocities (astrometric quality cuts are detailed in Appendix~\ref{appen:cuts}). We select Sun-like main-sequence stars following the same procedure and criteria as done for the hot Jupiter host sample in Section~\ref{subsect:sample-construct}.
After obtaining the stellar kinematics, we exclude from our field star sample the likely halo stars with a total Galactic rectangular velocity $v_{\rm tot} \equiv \sqrt{U^2+V^2+W^2}>180 \ {\rm km} \ {\rm s}^{-1}$ \citep[e.g.,][]{nissen2010two,bensby2014exploring,yan2019chemical}.

\added{Our field star sample is a close match to the HJ host sample, with the contamination from undetected HJ hosts, moving-group members, or young cluster members too small to affect our conclusion. The undetected HJ hosts are negligible in the field star sample, as the hot Jupiter occurrence rate is typically $\lesssim 1\%$ \citep[e.g.,][]{Gan:2023}. Within the solar neighborhood, young cluster or moving-group members may contribute up to $\sim10\%$ of all field stars \citep{Gaia2021}, but with our exclusion of very early/late type stars, this fraction is expected to be even lower in our selected sample of field stars. The contamination from thick disk stars at the solar neighborhood is also expected to be as low as $<10\%$.}

\subsection{Kinematic Age Comparison}\label{subsect:HS19-res}

\begin{deluxetable}{lc}
\tablecaption{Matching criteria for the field star comparison sample \label{tab:matching}}
\tablehead{
\colhead{Property} & \colhead{Condition}
}
\startdata
    (\text{RA}, \text{Dec}) & $\theta({\rm HJ}, {\rm FS})$\tablenotemark{a} $\leq 10^\circ$ \\
    Distance & $d_{\rm HJ}-100\ {\rm pc}\leq d_{\rm FS} \leq d_{\rm HJ}+50\ {\rm pc}$ \\
    $G$ magnitude & $G_{\rm HJ}-1\leq G_{\rm FS} \leq G_{\rm HJ}+0.1$ \\
    BP-RP color & $|(\rm{BP-RP})_{0,\rm FS}-(\rm{BP-RP})_{0,\rm HJ}|\leq 0.2$\\
\enddata
\tablecomments{The subscript ``HJ" denotes physical properties for hot Jupiter hosts, and ``FS" denotes those for Gaia field stars.}
\tablenotetext{a}{$\theta({\rm HJ}, {\rm FS})$ means the angular separation between the HJ host and the field star in degree.}
\end{deluxetable}

\begin{figure*}[]
    \centering
    \includegraphics[width=\linewidth]{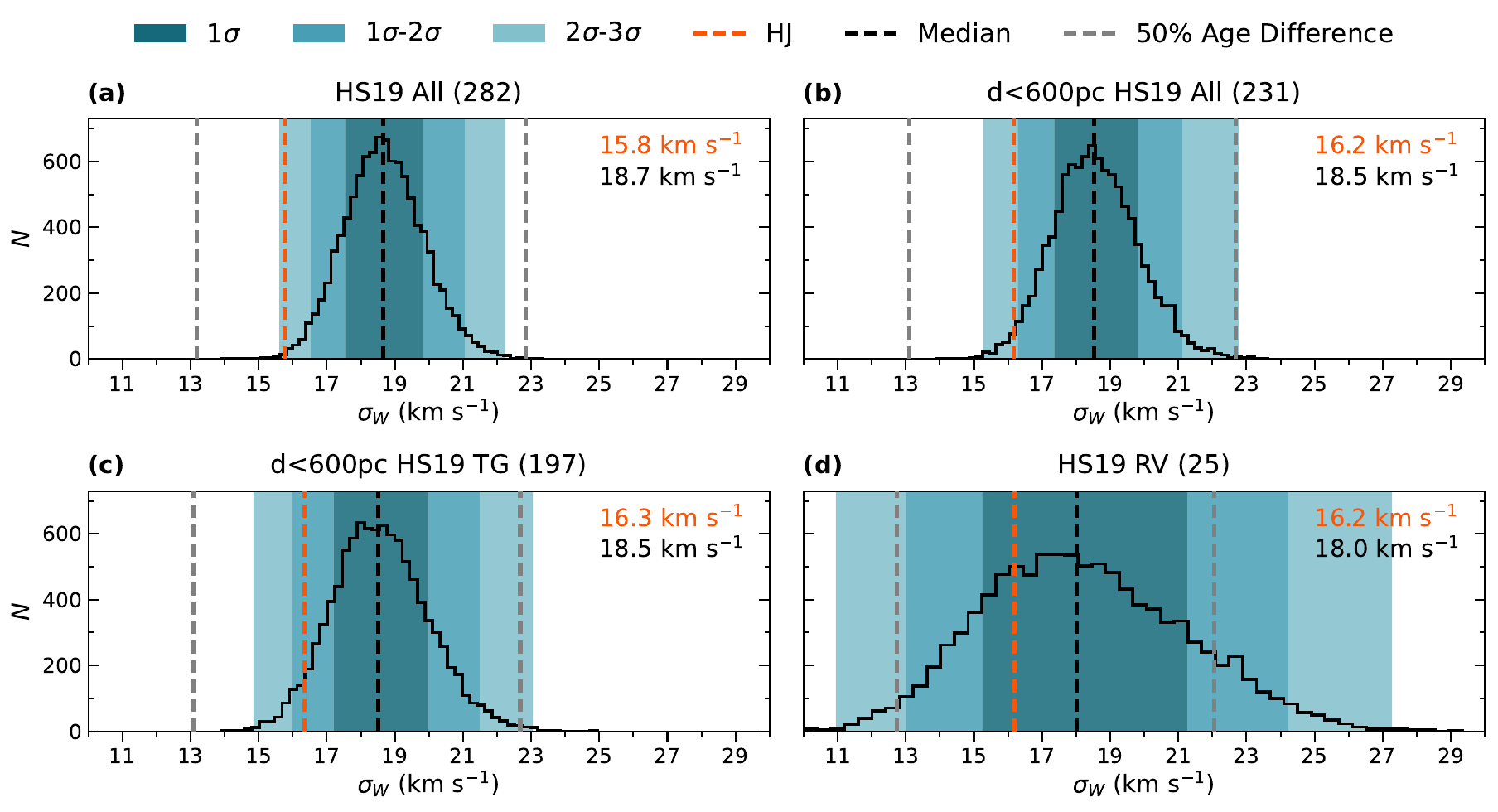}
    \caption{Vertical velocity dispersion ($\sigma_W$) of hot Jupiter host samples (orange dashed lines) versus those of matched field star control samples (black histograms) for the updated \citetalias{hamer2019hot} sample. The shaded regions and dashed black line indicate the 68\%-95\%-99.7\% confidence intervals and median of the field star histograms. The gray dashed lines indicate the $\sigma_W$ threshold at which the relative age difference from the field-star median reaches 50\%. The $\sigma_W$ values for the HJ sample and the field star median are given in the upper right corners in orange and black respectively. Panels (a) and (b): the whole updated \citetalias{hamer2019hot} sample, without and with a distance cut of $d<600$ pc. Panels (c) and (d): ground-based transit (TG) and radial velocity (RV) hosts in the updated \citetalias{hamer2019hot} sample with $d<600$ pc.}
    \label{fig:HS19}
\end{figure*}

\begin{figure*}
    \includegraphics[width=\linewidth]{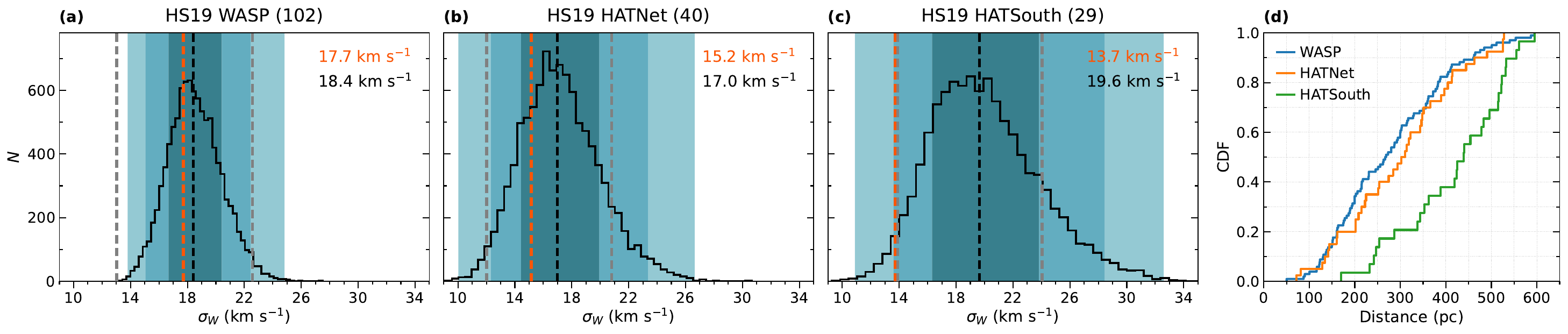}
    \caption{Panel (a)-(c): Vertical velocity dispersions ($\sigma_W$) of hot Jupiter hosts from three ground-based transit surveys (WASP, HATNet, and HATSouth) in the updated \citetalias{hamer2019hot} sample with $d<600$ pc versus those of matched field star samples. The legends are the same as those in Fig.~\ref{fig:HS19}. Panel (d): Cumulative distribution functions of host star distances for the three hot Jupiter samples shown in panels (a)-(c).}
    \label{fig:tg_subset}
\end{figure*}

\begin{figure}
    \includegraphics[width=\linewidth]{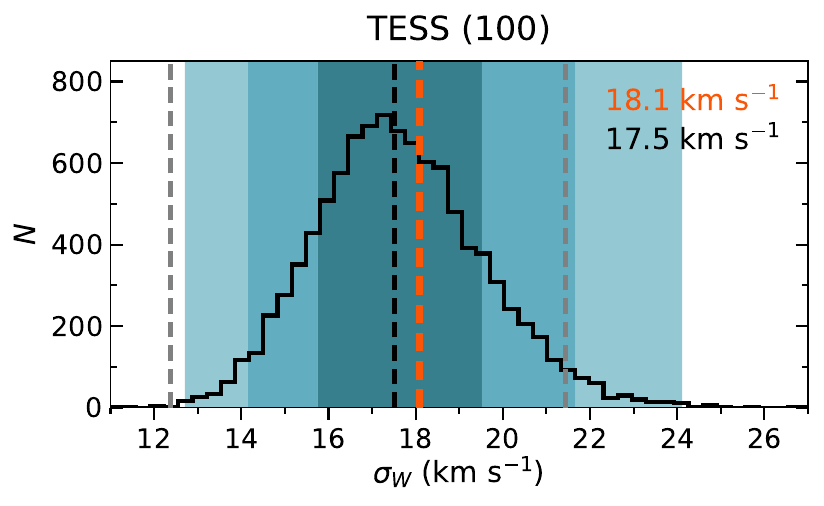}
    \caption{The vertical velocity dispersion ($\sigma_W$) for TESS hot Jupiter hosts with $d<600$ pc versus those of the matched field star sample. The legends are the same as in Fig.~\ref{fig:HS19}.}
    \label{fig:tess}
\end{figure}

As explained in Section~\ref{sect:intro}, we use the vertical velocity dispersion ($\sigma_W$) as the more robust age indicator. The vertical velocity dispersion is calculated in the standard way
\begin{equation}
\sigma_W \equiv \sqrt{\frac{1}{N}\sum_{i=1}^{N}\left(W_i - \overline{W} \right)^2},
\end{equation}
where $W_i$ are the individual vertical velocities, $\overline{W}$ is their mean, and $N$ is the number of stars in the sample.

In order to take into account the heterogeneity and anisotropy of the hot Jupiter sample as stated in Section~\ref{sect:intro}, we construct a control sample of field stars for each hot Jupiter host by matching them in right ascension, declination, distance, Gaia apparent $G$ magnitude, and intrinsic Gaia ${\rm BP-RP}$ color. Specifically, the HJ host and the field stars are matched within an angular separation of $10^{\circ}$, a distance difference of $[-100, +50]$ pc, a $G$-magnitude difference of $[-1, +0.1]$ mag, and an intrinsic ${\rm BP-RP}$ color difference of $0.2$, as summarized in Table~\ref{tab:matching}.
The asymmetric matching criteria for distance and $G$ magnitude are designed to balance the numbers of bright, nearby stars and faint, distant ones. Our matching criteria are stricter than those used in \citetalias{hamer2019hot}, which only matched Galactic height and stellar color. Simultaneously matching equatorial coordinates, distance, apparent $G$ magnitude, and intrinsic color effectively confines the control field stars to the three-dimensional spatial vicinity and the color--magnitude diagram vicinity of the hot Jupiter hosts, which mimic very roughly the target selection preferences in hot Jupiter detections of various methods/facilities. On average, each HJ host in the updated \citetalias{hamer2019hot} sample has $\sim 1,000$ matched stars in the field star control sample.

To compare the relative ages of hot Jupiter hosts and field stars, 
we perform Monte Carlo simulations described in the following. In each iteration, we randomly draw one field star for each hot Jupiter host from its matched field star sample, and compute the $\sigma_W$ of the sample of randomly drawn field stars.
This process is repeated for 10,000 times with replacement to obtain a distribution of $\sigma_W$ values for the matched field stars, which could be directly compared to $\sigma_W$ of the HJ sample.

We first perform the Monte Carlo experiment on all HJ hosts in the updated \citetalias{hamer2019hot} sample, and the result is shown in panel (a) of Fig.~\ref{fig:HS19}, where the black-edged histogram is the $\sigma_W$ distribution of the field stars, and the orange dashed line is the $\sigma_W$ of the updated \citetalias{hamer2019hot} HJ sample. To more intuitively illustrate the connection between $\sigma_W$ and mean age, we mark with gray dashed lines in Fig.~\ref{fig:HS19} the $\sigma_W$ threshold at which the relative age difference from the field-star median reaches 50\%. The kinematic $\sigma_W$ difference is converted into mean stellar age by rewriting the AVR from $\sigma_W \propto T^\beta $ to
\begin{equation} \label{eqn:avr}
    \frac{T_{\rm HJ}}{T_{\rm FS}} \approx \left(\frac{\sigma_{W, {\rm HJ}}}{\sigma_{W, {\rm FS}}} \right)^2 ,
\end{equation}
with ``FS'' standing for field stars. In the above equation we have adopted $\beta\approx0.5$\citep[e.g.,][]{holmberg2009genevacopenhagen,chen2021planets}. \footnote{\added{Although the AVR is affected by spatial and sample-dependent systematics, the inferred relative age difference is not so sensitive to the assumed value of $\beta$ when the $\sigma_W$ difference is not large: Taking a representative value of $\sigma_{W, {\rm HJ}}/\sigma_{W, {\rm FS}} = 0.9$ as an example, the corresponding relative age differences for the (intentionally exaggerated) values of $\beta = 0.4, 0.5, 0.6$ are $23\%, 19\%,$ and $16\%$, respectively.}} 
According to this result, the vertical velocity dispersion in hot Jupiter hosts is smaller than the vertical velocity dispersion of matched field star sample at a statistically level of $\sim3\sigma$. This velocity difference is statistically less significant than that reported in \citetalias{hamer2019hot}, but the overall trend is consistent. In fact, if we use the total velocity dispersion $\sigma_{\rm tot}=\sqrt{\sigma_U^2+\sigma_V^2+\sigma_W^2}$, we can indeed obtain a more significant velocity difference between the hot Jupiter host and matched field star samples. This consistency shows that, even with our modified stellar sample and statistical method, we are able to reproduce well the overall result of \citetalias{hamer2019hot}.

We then investigate the impact of sample heterogeneity on the velocity dispersion difference.
In panel (b) of Fig.~\ref{fig:HS19}, we perform the Monte Carlo experiment on HJ hosts with distances less than 600~pc. This distance cut is necessary to account for the AVR variations across Galactic locations \citep[e.g.,][]{sharma2021fundamental, sun2025age, chen2021planets}, \added{and to exclude extremely distant HJs in the ``HS19 TS'' sample (see left panel of Fig.~\ref{fig:sky-dist}) which has complex kinematic behavior in the Kepler field (see Zhu \& Hu 2026 and references therein). After this distance cut, the $\sigma_W$ of HJ hosts increase from $15.8~\kms$ to $16.2~\kms$, while the $\sigma_W$ distribution of the matched field stars remains largely unchanged.} The statistical difference in $\sigma_W$ between hot Jupiter host and the matched field star sample is reduced to $\sim2\sigma$.
In panels (c) and (d) of Fig.~\ref{fig:HS19}, we further separate the remaining HJ hosts by their detection method. We exclude the ``HS19 TS'' sample from now on \added{as discussed above}. The $\sigma_W$ values of HJ hosts discovered by ground-based transit and radial velocity lie within the $2\sigma$ and $1\sigma$ ranges of the $\sigma_W$ distributions derived from their corresponding field star samples, respectively. Neither subsample shows statistically significant difference in $\sigma_W$.

\added{To ensure that the matching procedure itself does not introduce unintended biases, we perform an independent validation to estimate the intrinsic $\sigma_W$ distribution of the field star control sample. We randomly draw with replacement 200 field stars with $G$ mag brighter than 13.5 mag and distance smaller than 600 pc, calculate their $\sigma_W$, and repeat this processes for 10,000 times. The intrinsic $\sigma_W$ distribution of the field star control sample has a median of $19.45~\kms$ and 1$\sigma$ range of $18.02-20.98~\kms$, which is statistically consistent with that obtained from the matched-control sample.}

Is the reduced statistical difference caused by simply the reduced sample size? To show that this is not the case, we further identify in the ``HS19 TG'' sample with $d<600\,$pc hot Jupiters from three leading ground-based transit surveys, namely
WASP \citep{pollacco2006wasp}, HATNet \citep{bakos2004wide},and HATSouth \citep{bakos2013hatsouth}, each contributing 102, 40, and 29 hot Jupiter detections, respectively. The vertical velocity dispersions for these three subsamples are $17.7~\kms$, $15.2~\kms$, and $13.7~\kms$, respectively, and the differences are partly due to the anisotropic distributions of hot Jupiter hosts from individual subsamples.
As shown in the left three panels of Fig.~\ref{fig:tg_subset}, the $\sigma_W$ for hot Jupiters discovered by WASP and HATNet are statistically indistinguishable ($<1\sigma$) from the $\sigma_W$ distributions of the matched field stars, and only in the small sample of 29 hot Jupiters discovered by HATSouth is the measured $\sigma_W$ statistically smaller (by $\sim2\sigma$) than the typical $\sigma_W$ of the matched field stars. The $\sigma_W$ difference in HATSouth subsample is a consequence of two factors: $\sigma_W$ of HATSouth hot Jupiter hosts is small, and the $\sigma_W$ of its matched field stars is large, compared to those of either WASP or HATNet subsamples. The former can be attributed to the small sample size of HATSouth hot Jupiters: a random draw of 29 hot Jupiters out of the 102 WASP hot Jupiters yields a value of $\sigma_W$ whose 1-$\sigma$ range is 12.9--21.0 $\kms$. The relatively large $\sigma_W$ of HATSouth field stars, $19.6~\kms$, is probably due to the larger contamination of thick disk stars. A large fraction of HATSouth hot Jupiter hosts are located at high Galactic latitudes, and with a further distance, the contamination from thick disk stars can no longer be negligible.
Therefore, the reduced statistical difference is not caused by the reduced sample size.

Excluding the peculiar but small sample of 29 hot Jupiters from HATSouth, the $\sigma_W$ of the ``HS19 TG'' sample with $d<600\,$pc cut becomes $16.7~\kms$, and $\sigma_W$ of its matched field stars is $18.1~\kms$. The statistical difference is only at the marginal level of $1\sigma$. The typical age of hot Jupiter hosts is different from the typical age of the field stars by $\lesssim 15\%$ according to Eq.~\ref{eqn:avr}.

\section{The TESS Sample}\label{sect:tess}

Since \citetalias{hamer2019hot}, TESS mission has been discovering many more hot Jupiters, and it offers to date the most homogeneous hot Jupiter sample available, which we now use to verify our result from the previous section.

We construct the TESS hot Jupiter sample following the same procedure as in Section~\ref{subsect:sample-construct}. 
This results in a sample of 105 hot Jupiter hosts, whose spatial distribution and planetary parameters are also shown in Figures~\ref{fig:sky-dist} and \ref{fig:m-p}, respectively.  The identifiers of the TESS HJ sample is also listed in Appendix~\ref{sect:target}.
The distance distribution of TESS HJ hosts resembles that of HJ hosts discovered by ground-based transit in the \citetalias{hamer2019hot} sample, but the former contains fewer distant ($d>600$ pc) systems. Because TESS is an all-sky survey, its HJ hosts are distributed uniformly across the celestial sphere. After imposing the same distance cut ($d<600\,$pc) to the TESS sample, we are left with 100 HJ hosts.

We compare the vertical velocity dispersion $\sigma_W$ of the TESS HJ sample to the distribution of $\sigma_W$ of the matched field stars, which is obtained in the same way as detailed in Section~\ref{subsect:HS19-res}. As shown in Fig.~\ref{fig:tess}, the TESS HJ sample has $\sigma_W = 18.1$ km s$^{-1}$, which is very close ($<1\sigma$) to the median value of the field star sample ($\sigma_W = 17.5$ km s$^{-1}$). This $\sigma_W$ difference could be translated into a difference between the typical age of TESS HJ hosts and field stars of $\lesssim 7\%$ through Eq.~\ref{eqn:avr}. This result confirms and further strengthens our finding in Section~\ref{subsect:HS19-res}, namely that the vertical velocity dispersion and thus the age of the HJ sample are statistically similar to that of their field star counterparts.

\added{While the TESS HJ sample is to date the most homogeneous sample available, it may also be affected by selection effects, such as the follow-up incompleteness and RV confirmation biases, although it is unclear how such selection effects affect the age of HJ hosts at the population level. We leave it to some future studies that make use of a more uniform HJ sample, such as the TESS Grand Unified Hot Jupiter Survey \citep{Yee2022}.}

\begin{figure}
    \centering
    \includegraphics[width=\linewidth]{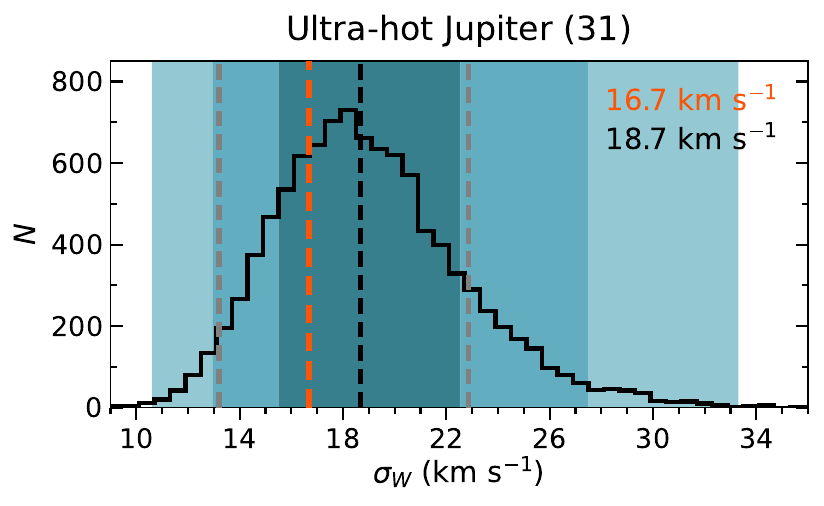}
    \caption{The vertical velocity dispersion ($\sigma_W$) for ultra-hot Jupiter sample versus those of the matched field star sample. The legends are the same as in Fig.~\ref{fig:HS19}.}
    \label{fig:shortP}
\end{figure}

\section{Ultra-hot Jupiter Sample}\label{sect:ultra-hot}

Motivated by the theoretical arguments that tides can only be strong enough to drive tidal disruption of Jupiter-mass planets with $P\lesssim 2\,$days for main sequence stars \citep[e.g.,][]{weinberg2024orbital}, we construct an ultra-hot Jupiter sample that is theoretically more vulnerable to tidal disruption, and see if this sample is statistically younger than their field star counterparts.

We define an ultra-hot Jupiter (UHJ) as a planet with orbital period $P<2\,$d and mass in the range $0.3<M_{\rm p} (\sin i)<13 \mj$.
We restrict to planets discovered by ground-based transit and TESS surveys to minimize the observational bias between different discovery methods.
We also exclude planets with distances larger than 600 pc to mitigate the effect of the AVR variations across Galactic locations.
Furthermore, we restrict our UHJ sample to stars with $0.8\leq (\rm{BP-RP})_0\leq1.5$, in order to exclude the F-type stars that are probably evolved but not excluded by the fairly generous definition of stellar main sequence. Starting from the NEA table and following a similar selection process as in Section~\ref{subsect:sample-construct}, we obtain a sample of 31 UHJ hosts, whose mass and period distributions are also shown by red-edged points in Fig.~\ref{fig:m-p}. Out of this UHJ sample, 19 are already in the \citetalias{hamer2019hot} HJ sample, and 12 are discovered afterwards by either ground-based transit (9) or TESS (3).

We compute the vertical velocity dispersion $\sigma_W$ of our UHJ sample and compare it with the distribution of $\sigma_W$ from the matched field stars, again following the same procedure as in Section~\ref{subsect:HS19-res}. The result is illustrated in Fig.~\ref{fig:shortP}. The $\sigma_W$ of the ultra-hot Jupiter sample ($16.7~\kms$) is within the $1\sigma$ range of the median $\sigma_W$ of the matched field stars ($18.7~\kms$), which can be translated into a typical age difference $\lesssim 20\%$. This means that even for the ultra-hot Jupiters, which are theoretically most vulnerable to tidal disruption, their kinematic ages are still statistically similar to, and certainly not significantly younger than, the ages of their field star counterparts.

There are potential caveats to the above result. First, the current UHJ sample is still small. Depending on the expected tidal decay rate, we may not be able to detect the signal even if there is one. On the other hand, the above analysis has not taken into account the sample heterogeneity, which may further reduce the difference in $\sigma_W$, as we have shown in Section~\ref{sect:HS19}. Another complication is on the physical mechanism that drives the tidal decay at such short periods. Compared to the nonlinear damping \citep{essick2015orbital,weinberg2024orbital} or wave breaking \citep{barker2010internal,barker2020tidal}, which predicts rapid tidal decay and disruption at $P\lesssim2$ days, the orbital evolution of ultra-hot Jupiters may instead be dominated by tidal resonance locking \citep{ma2021orbital}, which predicts that the tidal dissipation is weak at such short orbital periods.

\section{Hot Jupiters vs.\ Warm/Cold Jupiters} \label{sec:wcj}

\begin{figure}
    \centering
    \includegraphics[width=\linewidth]{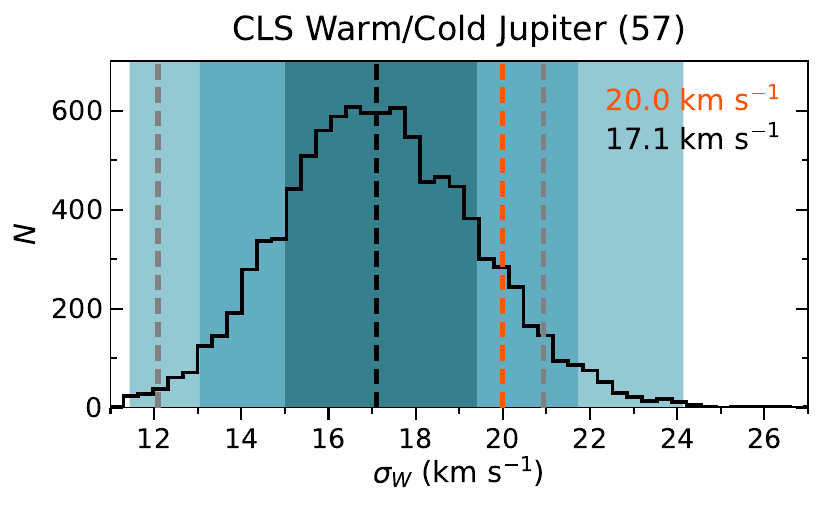}
    \caption{The vertical velocity dispersion ($\sigma_W$) for the cold Jupiter sample versus those of the matched field star sample. The legends are the same as in Fig.~\ref{fig:HS19}.}
    \label{fig:cj}
\end{figure}

Following \citetalias{hamer2019hot}, several studies have also used the warm/cold Jupiter hosts as reference and showed that they were kinematically older than the hot Jupiter hosts \citep{tejadaarevalo2021further, Mustill:2022, chen2023evolution, banerjee2024hoststar}. As warm/cold Jupiters have so long orbital periods that are not expected to be affected by stellar tides, these results were also taken as evidence supporting that hot Jupiters underwent significant tidal decay during stellar main sequence.

The majority of the warm/cold Jupiters are found by RV. As so, we take the warm/cold Jupiter systems from the California Legacy Survey \citep[CLS,][]{Rosenthal2021} and select those around Sun-like main-sequence hosts following the procedure and criteria in \citet{LiZhu2026}, in order to reduce the impact of selection biases in the original CLS sample \citep{Zhu2022}. This yields a sample of 57 warm/cold Jupiter systems with $P>10$~days. 
The vertical velocity dispersion of this RV warm/cold Jupiter sample is $20.0~\kms$, which is larger than the vertical velocity dispersion of the full hot Jupiter sample ($\sim16~\kms$, Fig.~\ref{fig:HS19}). This is broadly consistent with the result of earlier studies \citep{hamer2019hot, tejadaarevalo2021further, Mustill:2022, chen2023evolution,banerjee2024hoststar}. However, the RV surveys, which discovered the majority of the warm/cold Jupiters, and the transit surveys, which discovered the majority of the hot Jupiters, have very different target selection strategies, and their planet hosts differ in their brightness and 3D spatial distribution (Fig.~\ref{fig:sky-dist}). It is inappropriate to compare the stellar kinematics from these different samples directly, as has been demonstrated earlier in this work.

Given that $\sigma_W$ of HJ hosts is statistically similar to that of their matched field stars, we are then interested in knowing how $\sigma_W$ of warm/cold Jupiter hosts is compared to their matched field stars.
We construct the vertical velocity dispersion distribution of matched field stars in the same manner as in Section~\ref{subsect:HS19-res}, except for relaxing the matching criteria on the angular separation to $<20^{\circ}$, Gaia apparent $G$ magnitude difference to $[-1, +0.5]$ mag, and the distance difference to 100 pc, since the RV targets with warm/cold Jupiters are extremely bright and close. On average, each WJ/CJ host has $\sim 100$ matched stars in the field star control sample.
The resulting distribution, as shown in Fig.~\ref{fig:cj}, suggests that the vertical velocity dispersion of these RV warm/cold Jupiters is generally consistent with, although slightly larger (by $1.3\sigma$) than, that of the field stars.

Furthermore, $\sigma_W$ of the RV hot Jupiter sample is consistent with that of the matched field stars more than $\sigma_W$ of the warm/cold Jupiter sample is to its corresponding field stars. This points to an alternative explanation for the kinematic age difference between hot Jupiter hosts and warm/cold Jupiter hosts: while hot Jupiter hosts have consistent kinematics to the field stars, warm/cold Jupiter hosts are somewhat kinematically hotter (and thus older) than the field stars. This is a fairly plausible explanation, given that RV surveys preferentially select quieter and more slowly-rotating (thus older) stars for RV monitoring \citep[e.g.,][]{Santos2000, Wright2004}. Compared to hot Jupiters, warm/cold Jupiters are much harder to detect via RV and thus more significantly affected by the selection bias. 
This may explain, at least partly, the finding of \citet{miyazaki2023evidence} that the isochrone age of cold Jupiter hosts was on average larger than that of the hot Jupiter hosts in the CLS sample.
\footnote{Note that \citet{miyazaki2023evidence} had only nine hot Jupiters in the analysis and included evolved stars in their sample.}
We defer a systematic study of the hot Jupiter occurrence rate evolution to some future study.

\section{Summary}\label{sect:summary}

We revisit the study by \citetalias{hamer2019hot} to reassess whether main-sequence stars hosting hot Jupiters are kinematically younger than their field star counterparts. Our findings can be summarized as following:
\begin{itemize}
    \item After accounting for heterogeneity and anisotropy in the \citetalias{hamer2019hot} hot Jupiter sample, we find that the vertical velocity dispersion ($\sigma_W$) of the hot Jupiter hosts is not statistically different ($<1\sigma$) from that of the matched field stars (Fig.~\ref{fig:HS19}).
    \item As an independent validation, we analyze a new sample of hot Jupiters discovered by TESS and confirm that their $\sigma_{W}$ is also statistically indistinguishable from matched field stars (Fig.~\ref{fig:tess}). 
    \item We examine ultra-hot Jupiters with periods \(P<2\)~days, which are more vulnerable to tidal disruption, and find no statistically significant difference in $\sigma_W$ from matched field stars, either (Fig.~\ref{fig:shortP}).
    \item We find that $\sigma_W$ of RV warm/cold Jupiters, which is larger than that of hot Jupiters, is also generally consistent with that of the matched field stars. The difference in ages of hot Jupiters and warm/cold Jupiters may be caused by the selection bias in RV surveys.
\end{itemize}

\added{
Our results suggest that the systematics in the age--velocity dispersion relation is at the level of several $\kms$, and after accounting for the systematics, there is no statistically strong evidence from the stellar kinematics that a large fraction of hot Jupiters around Sun-like stars are tidally destroyed during the stellar main sequence.
Therefore, one should be cautious in applying the age--velocity dispersion relation to statistical studies of exoplanet hosts, especially if the expected age difference is relatively small ($\lesssim 20\%$).
}

\begin{acknowledgments}
We would like to thank Subo Dong, Kento Masuda, and Josh Winn for discussions and comments on an earlier version of the manuscript.
\added{We thank the anonymous reviewer for comments on the manuscript.}
Work by Q.H.\ and W.Z.\ was supported by the National Natural Science Foundation of China (grant Nos.\ 12173021 and 12133005). 
Y.H. acknowledges support from the National Natural Science Foundation of China (grant No. 12422303).
\end{acknowledgments}

\software{\texttt{numpy} \citep{harris2020array},
          \texttt{matplotlib} \citep{hunter2007matplotlib},
          \texttt{PyAstronomy} \citep{pyastronomy}, 
          \texttt{astropy} \citep{2022ApJ...935..167A},
          \texttt{dustmaps} \citep{green2018dustmaps}
          }

\appendix

\section{Astrometric Quality Cuts}\label{appen:cuts}
We follow the astrometric quality cuts used in \citetalias{hamer2019hot} to select our hot Jupiter and field star sample. We replace their cut \texttt{astrometric\_excess\_noise\_sig}$<2$ by \texttt{RUWE} $< 1.4$, since \texttt{RUWE} is a better indicator of goodness of fitting widely used in Gaia DR3 data. Besides, the $-0.23<\texttt{mean\_varpi\_factor}<0.36$ cut is removed, since Gaia DR3 dataset does not provide the \texttt{mean\_varpi\_factor}. To the hot Jupiter host sample, we only impose cuts 1-3, since the reflex motion of the host star caused by the giant planet can result in excess noise in the astrometic fitting. The astrometric quality cuts are as follows,
\begin{enumerate}[itemsep=0pt]
    \item $\texttt{parallax\_over\_error}>10$
    \item $\texttt{visibility\_periods\_used}>8$
    \item $u<1.2 \times \texttt{Max}(1,\exp(-0.2(\texttt{phot\_g\_mean\_mag}-19.5)))$\footnote{A parentheses is missing in Cut 4 in the appendix of \citetalias{hamer2019hot}.}
    \item $\texttt{astrometric\_gof\_al} <3$
    \item $\texttt{RUWE} < 1.4$
    \item $\texttt{rv\_nb\_transits} >5$
    \item $1.0+0.0015 \times \texttt{bp\_rp}^2 < \texttt{phot\_bp\_rp\_excess\_factor} < 1.3 + 0.06 \times \texttt{bp\_rp}^2$
    \item $\texttt{phot\_bp\_mean\_flux\_over\_error} > 10$
    \item $\texttt{phot\_rp\_mean\_flux\_over\_error} > 10$
\end{enumerate}

\section{Target Identifiers}\label{sect:target}
\begin{table}[]
    \centering
    \caption{The Updated \citetalias{hamer2019hot} Sample}
    \label{tab:2019list}
    \begin{tabular}{ll}
    \hline\hline
    NASA Exoplanet Archive Name  &  Gaia DR3 ID\\
    \hline
    51 Peg   & Gaia DR3 2835207319109249920 \\
    BD-10 3166& Gaia DR3 3758629479636689536 \\
    CoRoT-1   & Gaia DR3 3105507886130792448 \\
    CoRoT-13  & Gaia DR3 3101974231859723776 \\
    CoRoT-18  & Gaia DR3 3120058170178779392 \\
    CoRoT-2   & Gaia DR3 4287820848378092672 \\
    CoRoT-20  & Gaia DR3 3120122805143587328 \\
    CoRoT-25  & Gaia DR3 4285689616891369600 \\
    CoRoT-27  & Gaia DR3 4284721191974346752 \\
    CoRoT-3   & Gaia DR3 4263427117767106816 \\
    ... & ... \\
    \hline\hline
    \end{tabular}
    \tablecomments{Table~\ref{tab:2019list} is organized in the alphabet order by the NASA Exoplanet Archive name, and is published in its entirety in a machine-readable format.}
\end{table}

\begin{table}[]
    \centering
    \caption{The TESS HJ Sample}
    \label{tab:tesslist}
    \begin{tabular}{ll}
    \hline\hline
    NASA Exoplanet Archive Name  &  Gaia DR3 ID\\
    \hline
    HD 2685 & Gaia DR3 4684205720883329920 \\
    HD 63433& Gaia DR3 875071278432954240 \\
    NGTS-31 & Gaia DR3 4819647205326045824 \\
    TOI-1107& Gaia DR3 5192364501633274880 \\
    TOI-1194& Gaia DR3 1074139748425931008 \\
    TOI-1199& Gaia DR3 861975270310252416 \\
    TOI-1268& Gaia DR3 1675923009431370880 \\
    TOI-1288& Gaia DR3 2245652826430109184 \\
    TOI-1294& Gaia DR3 1695343683312903168 \\
    TOI-1295& Gaia DR3 1636690613486307840 \\
    ... & ... \\
    \hline\hline
    \end{tabular}
    \tablecomments{Table~\ref{tab:tesslist} is organized in the alphabet order by the NASA Exoplanet Archive name, and is published in its entirety in a machine-readable format.}
\end{table}

In Table~\ref{tab:2019list} and Table~\ref{tab:tesslist}, we list the NASA Exoplanet Archive Names and the Gaia DR3 IDs for the updated \citetalias{hamer2019hot} sample in Section~\ref{sect:HS19} and the TESS HJ sample in Section~\ref{sect:tess}. Both tables are published online in their entirety in a machine-readable format.

\bibliography{ref}{}
\bibliographystyle{aasjournalv7}

\end{CJK*}
\end{document}